\documentclass[12pt]{article}
\usepackage{graphicx}
\usepackage{amssymb}
\usepackage{epstopdf}
\usepackage{natbib}
\newcommand{\be}{\begin{equation}} \newcommand{\ee}{\end{equation}}
\newcommand{\ba}{\begin{eqnarray}} \newcommand{\ea}{\end{eqnarray}}
\newcommand{\nn}{\nonumber} \renewcommand{\bf}{\textbf}
\newcommand{\ra}{\rightarrow}

\renewcommand{\a}{\alpha}

\input{epsf}
\begin{document}

\title{ Signals of Statistical Anisotropy in WMAP Foreground-Cleaned Maps}
\author{ Pramoda Kumar Samal$^{1}$, Rajib Saha$^{2}$, Pankaj~Jain$^{1}$
\\     John~P.~Ralston$^3$}
\maketitle

\begin{center}
{$^{1}$Department of Physics, Indian Institute of Technology, Kanpur, U.P, 208016, India \\ }
{$^{2}$Jet Propulsion Laboratory, M/S 169-327, 4800 Oak   Grove Drive, Pasadena,
CA 91109; California Institute of Technology, Pasadena, CA 91125\\ }
{$^{3}$Department of Physics \& Astronomy,
University of Kansas, Lawrence, KS 66045, USA\\}
\end{center}
emails: samal@iitk.ac.in, Rajib.Saha@jpl.nasa.gov, pkjain@iitk.ac.in, ralston@ku.edu

\newpage
\begin{abstract}
Recently a symmetry-based method to test for statistical isotropy of
the cosmic microwave background was developed. We apply
the method to template-cleaned 3-year and 5-year WMAP-$DA$
maps. We examine a wide range of angular multipoles from
$2 < l < 300$. The analysis detects statistically signicant signals of
anisotropy inconsistent with an isotropic CMB in some of the foreground
cleaned maps. We are unable to resolve whether the anomalies
have a cosmological, local astrophysical or instrumental
origin. Assuming the anisotropy arises due to residual
foreground contamination, we estimate the residual foreground power
in the maps. For the W band maps, we also find a highly improbable  degree
of isotropy we cannot explain.  We speculate that excess isotropy may be
 caused by faulty modeling of detector noise. 
\end{abstract}

\section{Introduction}
 The inflationary Big Bang model assumes that anisotopies of the cosmic microwave background ($CMB$) come from random isotropic perturbations in the early universe. However there are indications that cosmological observables may 
 not be isotropic.
The indications include  distributions of polarizations from radio galaxies 
(Birch 1982,  Kendall and Young 1984, Jain and Ralston 1999, Jain and Sarala 2006),  statistics of optical polarizations from quasars (Hutsem\'{e}kers 1998, Hutsem\'{e}kers and Lamy 2001, Jain {\it et al} 2004) and many studies of 
unpolarized $CMB$ data. The CMB studies indicate an alignment of the
low-$l$ multipoles (de Oliveira-Costa
{\it et al} 2004, Ralston and Jain 2004, Schwarz {\it et al} 2004) and a 
hemispherical anisotropy (Eriksen {\it et al} 2004). The indications
of violation of isotropy in CMB data has prompted subtantial activity
 with varying outcomes 
(Katz and Weeks 2004, Bielewicz {\it et al} 2004, Hansen {\it et al} 2004, 
Bielewicz {\it et al} 2005, Prunet {\it et al} 2005, Copi {\it et al} 2006,
de Oliveira-Costa and Tegmark 2006, Wiaux {\it et al} 2006, Bernui {\it et al} 
2006, Freeman {\it et al} 2006, Magueijo and  Sorkin 2007, Bernui {\it et al} 
2007, Copi {\it et al} 2007, Eriksen {\it et al} 2007b,
Helling {\it et al} 2007, Land and Magueijo 2007, Pullen and Kamionkowski 2007,
Lew 2008, Bernui 2008). Differences arise 
due to different tests being used by different authors  
(Efstathiou 2003, Hajian {\it et al} 2004,
Hajian and Souradeep 2006, Donoghue and  Donoghue 2005) and radio 
(Bietenholz and Kronberg 1984) data. Despite a measure 
of controversy, it is astonishing that
diverse data sets all indicate a common
axis of anisotropy, pointing roughly in the direction of the Virgo supercluster
(Ralston and Jain 2004).

The possible violation of statistical isotropy in CMB has lead to many
theoretical studies (Cline {\it et al} 2003, Contaldi
  {\it et al} 2003, Kesden {\it et al} 2003, Berera {\it et al} 2004,
    Armendariz-Picon 2004, Moffat 2005, Gordon {\it et al} 2005, 
  Land and Magueijo 2005,
    Vale 2005, Abramo {\it et al} 2006, Land and Magueijo 2006, Rakic {\it et al} 2006
Gumrukcuoglu {\it et al} 2006, Inoue and Silk 2006, Rodrigues 2008,
Naselsky {\it et al} 2007, Campanelli {\it et al} 2007,
Koivisto and Mota 2007, Boehmer and Mota 2008, Kahniashvili {\it et al} 
2008, Dimopoulos {\it et al} 2008 ). The generation and evolution of primordial
perturbations in an anisotropic universe has also been studied
(Koivisto and Mota 2006,  Battye  and  Moss 2006, Armendariz-Picon 2006, Pereira {\it et al} 2007, Gumrukcuoglu {\it et al} 2007) as well as the possibility
of anisotropic inflation 
(Hunt and Sarkar 2004, Buniy {\it et al} 2006, Donoghue {\it et al} 2007,
Yokoyama and Soda 2008, Kanno {\it et al} 2008,  Erickcek {\it et al} 2008). 
The possibility that foreground contamination can lead to alignment has  
been investigated (Gaztanaga {\it et al} 2003, Slosar and Seljak 2004). 
Alternatively it has been suggested that systematic and statistical errors
in the extracted CMB signal may lead to the observed anomalies 
(Liu and Li 2008).
There have also been some theoretical studies of the optical alignment effect 
(Jain {\it et al} 2002, Payez {\it et al} 2008, Hutsem\'{e}kers 
{\it et al} 2008).
It may be possible to explain the violation of isotropy in CMB and
radio polarizations due to some local effect. 
 However the alignment of optical
polarizations depends on redshift and hence cannot be attributed to 
a local effect (Jain {\it et al} 2002).

In a recent paper (Samal {\it et al} 2008) we  introduced a new method for 
testing isotropy of $CMB$ data. The method is based on
identifying  invariant relations between different multipoles. For each multipole $l\ge 2$ we
identify three rotationally invariant eigenvalues
of the {\it power matrix} $A_{ij}$, defined by
\begin{equation}
A_{ij} = {1\over l(l+1)} \sum_{m,m'} a^*_{lm}(J_iJ_j)_{mm'}a_{lm'}\, .
\end{equation}
Here $J_i$ ($i=1,2,3$) are the angular momentum operators in
representation $l$.
The sum of the eigenvalues is the
usual power $C_l$. The remaining independent combinations of
eigenvalues provide information about the
isotropy of the sample.

In an infinite isotropic sample all the eigenvalues of the power matrix would
be equal. Statistical anisotropies in $CMB$ data will certainly lead to statistical
fluctuations in the eigenvalues. We quantify the fluctuations by introducing the concept of {\it power
entropy}. The eigenvectors of the matrix $A_{ij}$ also contain additional
information. Their orientation should be random in truly isotropic data. We
define the ``principal" eigenvector as the one associated with the largest
eigenvalue. We then study the  {\it alignment entropy}, which tests
for alignment among different eigenvectors.

In Samal {\it et al} (2008), we studied the WMAP Interior Linear Combination
(ILC) data set and restricted attention to  the multipole region
$l\le 50$. In the present paper we study the individual 
foreground cleaned Differencing Assembly (DA) maps,
$Q1, \, Q2, \, V1, \, V2, \, W1, \, W2, \, W3, \, W4$, also prepared by the WMAP team.  We also extend the scope of analysis to the range
$2\le l\le 300$.  As far as we know these are the first such tests for high multipoles.  They illustrate the effectiveness of the method compared to others, such as Maxwell multipoles (Copi {\it et al} 2006, 2007, Weeks 2004, 
Katz and Weeks 
2004), which run into combinatoric problems at high $l$ (Dennis 2005).
 We do not use the ILC
map, since it is not expected to be reliable for the large $l$ range we
consider here. At large $l$ the WMAP team uses the bands
$V1, \, V2, \, W1,\, W2, \, W3, \, W4$ for
their final power extraction in the 3-year and 5-year analysis. The $Q1$
and $Q2$ bands were not used in WMAP power estimates since they were found to be significantly
contaminated by foreground effects.

Our motivation for the study is twofold. First, we
are interested in testing whether the anisotropies found in Samal {\it et al} 
(2008) continue to hold for a larger range of multipoles. Second, we wish
to test whether additional anomalies in this data may exist. Our tests are 
not intended to determine whether anomalies come from some physical effect, 
contamination due to foregrounds, or correlations of noise.

In next Section we briefly review the methodology.  In Section 3 we describe
how the methodology is applied to the WMAP data.
In Section 4 we give results for test of statistical isotropy using
 the power entropy. In Section 5 we test for alignment of different multipoles
with the quadrupole axis. In Section 6 we test for statistical isotropy
using the  alignment entropy.
We conclude in Section 7.

\section{Covariant Frames and Statistics Across Multipoles}
\label{frames}

The $CMB$ temperature fluctuation in each map is conventionally expanded in spherical harmonics \ba T(\hat n) = \sum_{lm} \, a_{lm}Y_{lm}(\hat n). \nn \ea  The usual power $C_{l} \sim \sum_{m} \, a_{lm}a_{lm}^{*}$ is rotationally invariant and has no information about anisotropy.   The angular orientation of each mode is probed by a unique orthonormal frame $e_k^{\alpha}(l)$ and rotationally invariant eigenvalues $\Lambda_{\a}(l)$.   These are obtained by diagonalizing the power tensor $A$, defined by \ba   A_{ij}= < \, a \, | J_{i} J_{j}   \, | \, a \,>, \nn \\
=   \sum_{\a}\, e_{i}^{\a }     ( \Lambda^{\a})^{2}   e_{j}^{\a *}. \nn \ea   Here $J_{i}$ is the rotation generator in representation $l$, and index $l$ is suppressed when obvious. 

Basic statistics derived from frames are the {\it power entropy} $S_{P}$ and the {\it alignment entropy} $S_{X}$. Entropy is defined as in quantum statistical mechanics. The power density matrix $\rho_{P} = A/tr(A)$, where $tr$ indicates the trace, is normalized, $tr(\rho_{p})=1$, to remove the power.   The power entropy $S_{P}$ for each
multipole is
\ba S_{P} &=&   -tr(\,   \rho_{P} \, log(\,   \rho_{P} \, ) \,).   \label{pow_entropy_eq}
\ea     Isotropy predicts the maximum entropy   
\ba S_{P} \ra log(3) \:\:\:\:\: (isotropy). \nn \ea Small values of $S_{p}$ indicates anisotropy.   Note these measures apply mode-by-mode.   The full range is $ 0   \leq   S_{P} \leq log(3)$, where $ S_{P} \ra 0$ for a ``pure state'' $\tilde \Lambda_{1}=1$ aligned along a single axis.

The alignment entropy $S_{X}$ is a measure of alignment of frame axes. Let $e^i(l)$ be the ``principal eigenvector'' of the power tensor, meaning the one with the largest eigenvalue.   Construct a $3\times 3$ matrix $X_{ij}$:
\begin{equation}
X_{ij} = \sum_{l=l_{min}}^{l_{max}}   \,   e^i(l)e^j(l) .
\end{equation}   This tensor probe effectively averages
over a range of multipole moments.
Normalize by computing $\tilde X = X/tr(X)$. 
The alignment entropy is \ba S_{X} =- tr( \tilde X log \tilde X). \nn \ea

\section{Application to WMAP data}
\label{sec:Application}

We use the
WMAP 3-year and 5-year data for our analysis. The WMAP team 
(Hinshaw {\it et al} 2003, 2007) provides
foreground-cleaned maps for the Q, V and W bands. 
The $V$ and $W$ bands are used for power spectrum estimation. The $Q$ 
band is not used since it is found to be significantly foreground contaminated.
The foreground removal method adopted by WMAP is incomplete in the galactic plane. This region is removed by using the $Kp2$ mask before power spectrum 
estimation. Applying Kp2 mask also eliminates emissions from the resolved 
point sources by removing
circular area of radii $0.6^\circ$ around the position of each of the sources.
There also exist other foreground cleaning procedures that may be interesting to
compare (Tegmark {\it et al} 2003, Saha {\it et al} 2006, Eriksen {\it et al}
2007a). Here we study only the foreground cleaned maps provided by the WMAP team. 

\subsection{Data Preparation}
We apply the $Kp2$ mask to all the individual foreground cleaned DA maps. 
The masked region is filled by a randomly generated CMB signal along with
simulated detector noise based on WMAP's noise characteristics appropriate to each of the 8 maps. 

Noise maps are generated as follows. Let $\sigma_0$ be the noise per observation of the detector under consideration. Let $N_{pix}$ denote the number of pixels in each $N_{side}= 512$ level resolution map, and $N_p$ be the effective number of observations at each pixel. Sample a Gaussian distribution with zero mean and unit variance $N_{pix}$ number of times. Multiply each Gaussian variable by $\sigma_0 /\sqrt{N_p}$ to form realistic detector noise maps.

\begin{figure}[h]
\centering
\includegraphics[scale=0.45,angle=0]{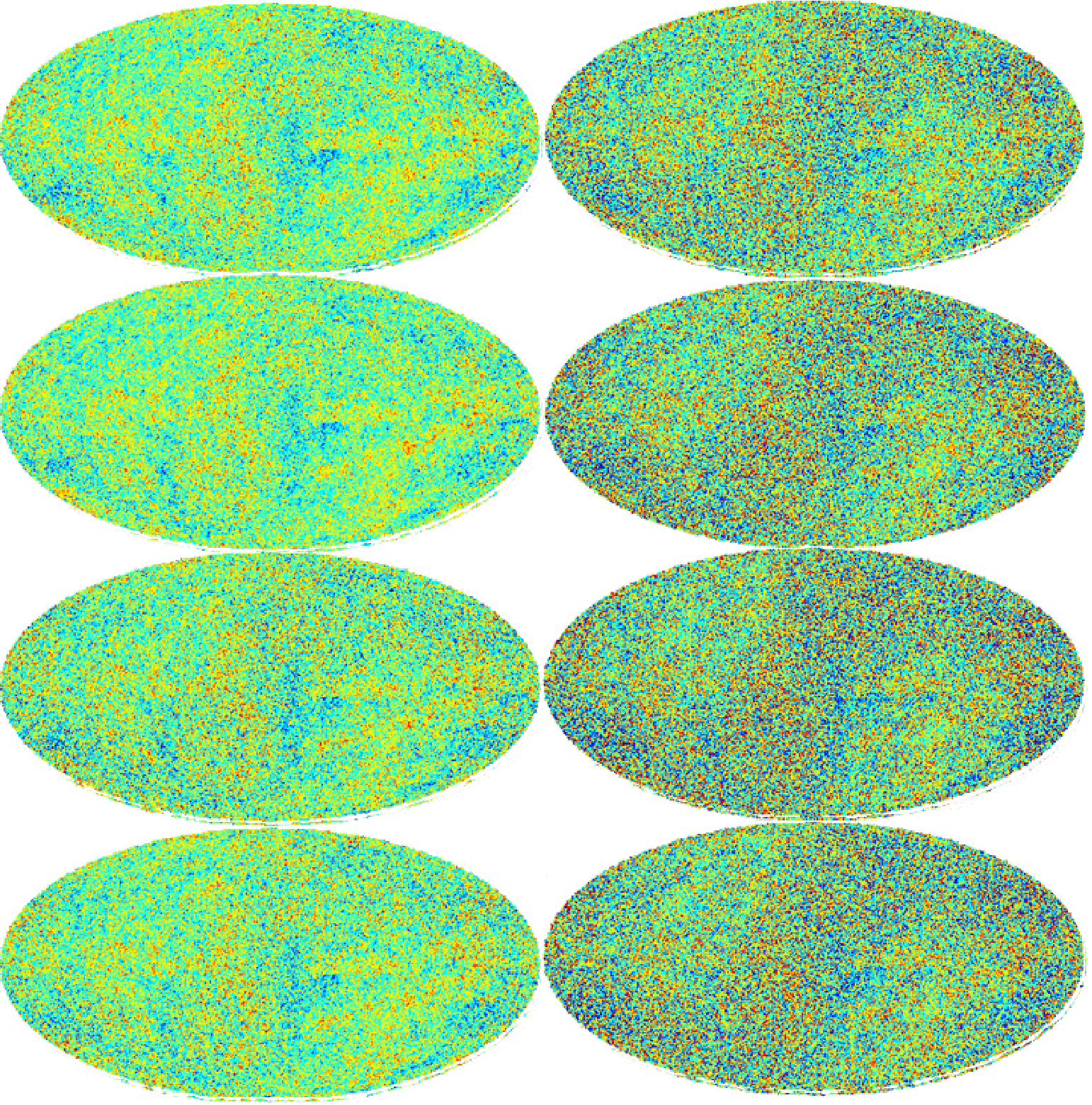}
\caption{ \small The 3-year maps after filling the $Kp2$ region with randomly generated $CMB$ signal
and detector noise appropriate for each band. 
From top to bottom 
the left panel shows $Q1$, $Q2$, $V1$ and $V2$ DA maps respectively while the
right panel shows $W1$, $W2$, $W3$ and $W4$ DA maps.}
\label{fig:8maps}
\end{figure}

Graphics of the $8$ maps used in our study are shown in Fig.~\ref{fig:8maps}. There is no visible signature of galactic
foreground contamination in the maps. Detector noise is evident in the W band
DA maps.

\subsection{Null Distributions}

Statistical baselines were developed from 10,000-run simulations of isotropic random $CMB$ power normalized to the data maps and including detector noise appropriate to each band. We set {\it preliminary} levels of statistical significance using $P$-values of 0.05 or less.  $P$ values are defined by the relative frequency for a statistic to occur with $P$ or less. The significance level of 
collections of $P$-values is estimated using the binomial distribution of 
``pass''and ``fail'' outcomes. The probability to encounter $k$ instances of 
passing defined by probability $p$ in $n$ trials is 
\ba  P_{bin}(k, \,p, \,n)= p^k(1-p)^{(n-k)}n!/(n-k)!k!. \nn \ea    
The binomial distribution is well-known, and we also verified the distribution describes $P$ values from the null simulations. In assessing many $P$-values we report the cumulative binomial probabilities 
\ba P_{bin}(k \geq k_* , \, p, \,n)= \sum^n_{k =k_*} \, P_{bin}(k, \,p, \,n). \nn \ea

\section{Power Entropy}
\label{power_entropy}

\begin{figure}[htb]
\includegraphics[scale= 0.55, angle = -90]{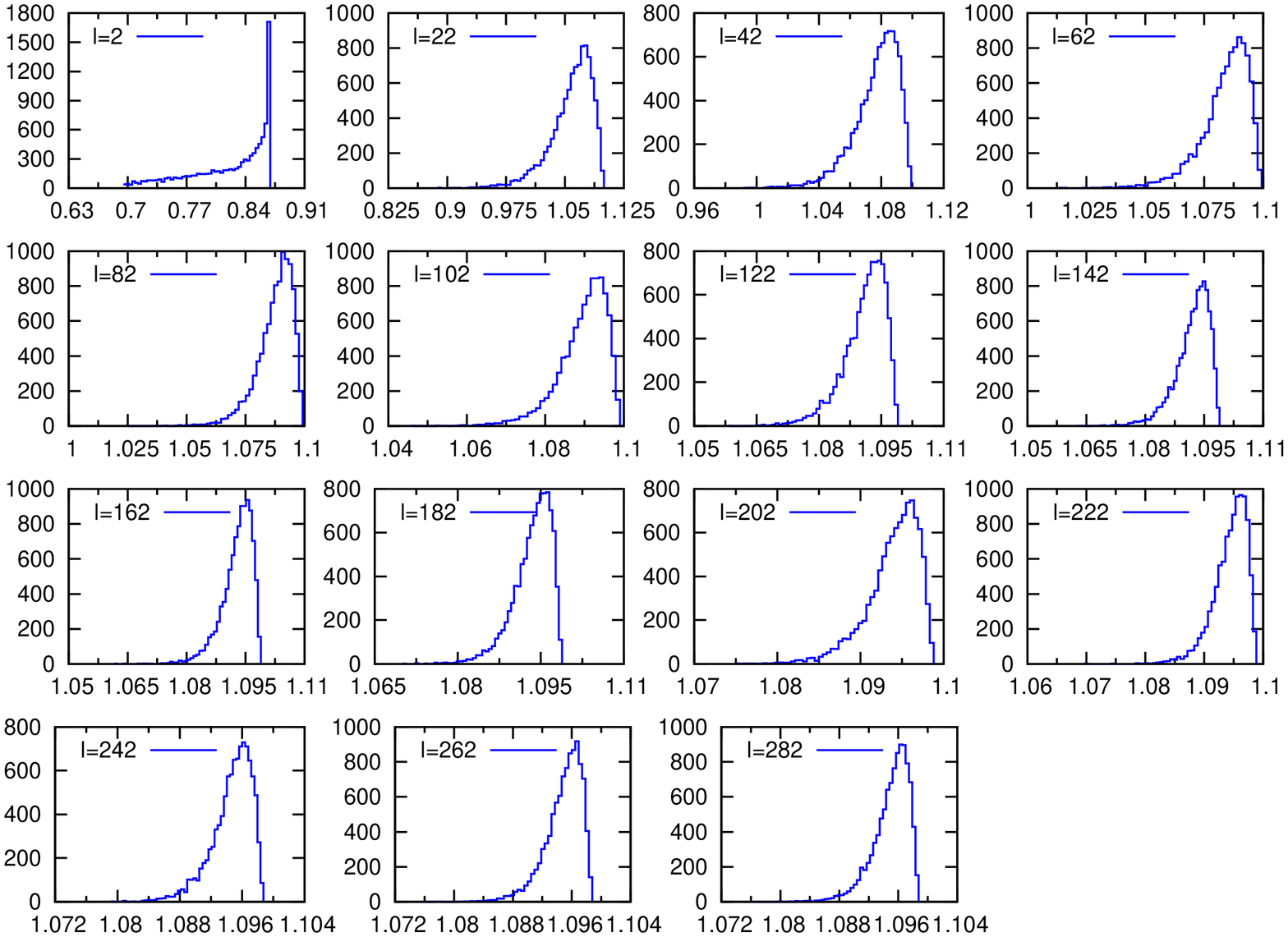}
\caption{ \small Histograms of the power entropy $S_{P}$ for multipole range
$2\le l \le 300$ at intervals of $20$ units using the WMAP 3-year data for
the $Q1$ map. }
\label{pow_entr_cmb_nse_hist}
\end{figure}

Fig.~\ref{pow_entr_cmb_nse_hist} shows the null distribution of power entropy for the $Q1$ map over the multipole range $2\le l \le 300$.   The distributions
of all the maps remains the same whether or not detector noise is added to the simulation.

\begin{figure}[h]
\includegraphics[width=7cm,height=14cm,angle=-90]{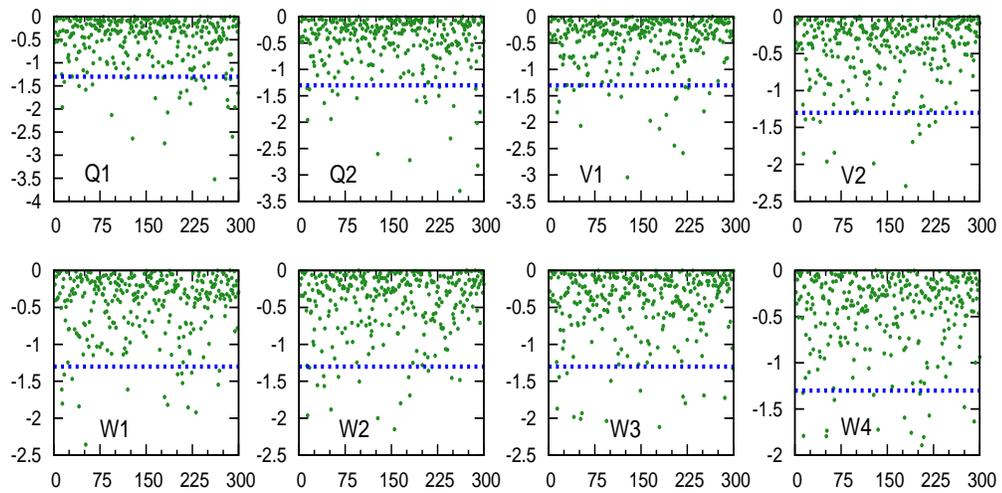}
\caption{ \small $log_{10}(P)$-values of the power entropy from the eight WMAP bands
for the range $2 \leq l \leq 300$ for the WMAP 3-year data.
The dashed horizontal line shows $P= 0.05$.}
\label{fig:Pval}
\end{figure}

Fig.~\ref{fig:Pval} shows $P$ values obtained from the WMAP data for the entire
range, $2\le l\le 300$, of multipole values considered. The horizontal
dashed line indicates $P =0.05$.
Violation of statistical isotropy is indicated for many
multipoles in all the bands.
Table~\ref{tab:tab1} (\ref{tab:tab1_5yr}) lists the 3-year (5-year)
multipoles for different bands with $P$-values potentially inconsistent 
with isotropy.

\begin{figure}[htb]
\includegraphics[scale=0.55,angle=-90]{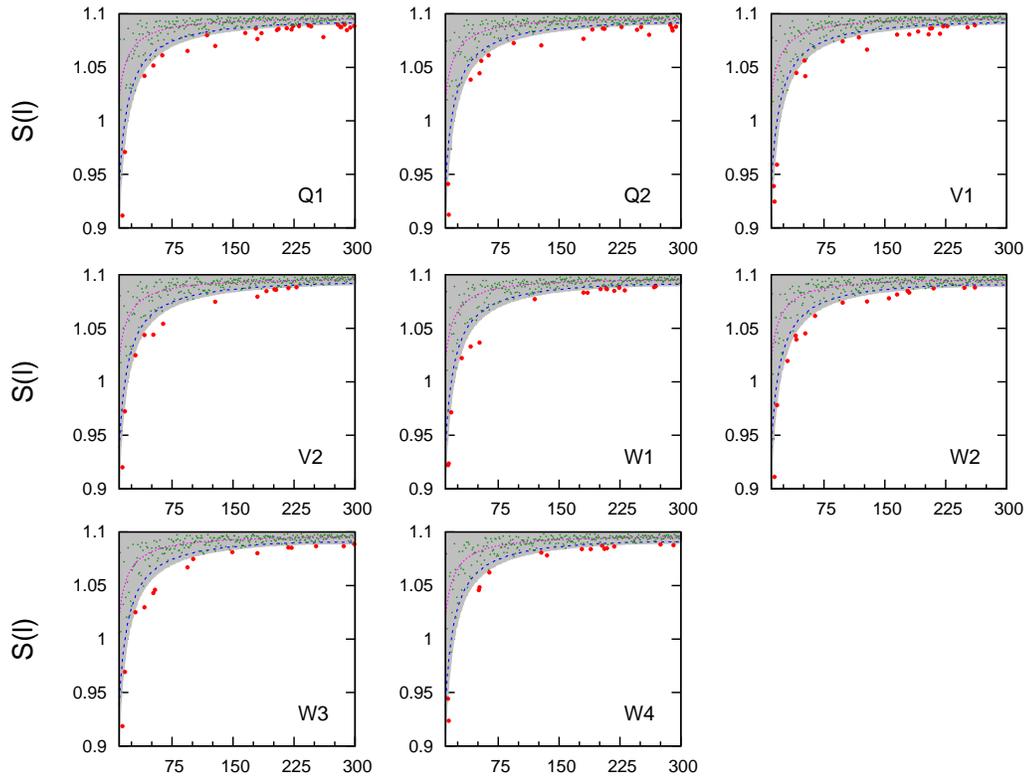}
\caption{ \small Distribution of the power entropy $S(l)$ showing the $95\%$ confidence level (grey band) for the WMAP 3-year data. Red points show multipoles potentially inconsistent with the isotropic prediction.}
\label{pow_entropy}
\end{figure}

\begin{table}
\scriptsize
\begin{center}
\begin{tabular}{c  l}
\hline
\hline
$Band$ &\ \ \ \ \ \ \ \ \ \ \ \ \ \ \ \ \ \ \ \ \ \ \ \ \ \  $Multipoles$ \\
\hline
\hline
& \\
Q1   & 14, 17, 41, 52, 63, 94, 118, 128, 165, 178, 180, 185, 204, 206, 216,   222, 224,   231, \\ & 243, 246, 261, 279, 280, 282, 283, 287, 290, 294, 299 \\ & \\ 
$Q2$   & 13, 14, 17, 41, 52, 54, 63, 94, 128, 180, 191, 204, 206, 227, 228, 246, 251,   261, 287, \\ & 289, 290, 294\\ & \\
$V1$   &   13, 14, 17, 41, 51, 52, 98, 118, 128, 165, 180, 191, 204, 206, 208, 218, 222, 227, 252, \\ &261 \\ & \\
$V2$   & 14, 17, 30, 41, 52, 64, 128, 180, 191, 201, 203, 218, 228 \\ & \\
$W1$   & 13, 14, 17, 30, 41, 52, 120, 180, 185, 201, 208, 209, 218, 224, 231, 267, 269 \\   & \\
$W2$   & 14, 17, 30, 40, 41, 52, 64, 98, 128, 155, 165, 178, 180, 210, 248, 261\\   & \\
$W3$   & 14, 17, 30, 41, 52, 54, 94, 101, 149, 180, 218, 222, 252, 286, 299 \\ & \\
$W4$   & 13, 14, 51, 52,   64, 128, 135, 178, 189, 203, 206, 209, 218, 275, 291 \\& \\ 
\hline   
\end{tabular}
\caption{ \small List of multipoles with $P < 0.05$ for power entropy
for the 3-year WMAP-DA maps.}
\label{tab:tab1}
\end{center}
\end{table}

\begin{table}
\scriptsize
\begin{center}
\begin{tabular}{c  l}
\hline
\hline
$Band$     & \ \ \ \ \ \ \ \ \ \ \ \ \ \ \ \ \ \ \ \ \ \ \ \  $Multipoles$     \\ 
\hline
\hline
& \\
$Q1$   &   14, 17, 41, 52, 94, 128, 135, 165, 177, 178, 180, 185, 191, 204, 206, 216, 218, 221,   \\ &
222, 225, 231, 261, 290, 294 \\ & \\
$Q2$   & 13, 14, 17, 41, 52, 54, 94, 128, 165, 170, 180, 191, 204, 206, 228, 246, 251, 261, \\ &
290, 294 \\ & \\
$V1$   &   13, 14, 17, 41, 52, 54, 64, 101, 128, 165, 180, 191, 204, 206, 218, 222, 231, 252, \\ &
290 \\ & \\
$V2$   &   14, 17, 30, 41, 52, 64, 94, 128, 161, 165, 180, 201, 204, 209, 218, 228 \\ & \\
$W1$   &   13, 14, 17, 30, 41, 52, 64, 120, 128, 139, 180, 185, 201, 204, 210, 218, 224, 228, 231, 269 \\   & \\
$W2$   &   13, 14, 30, 40, 41, 52, 98, 115, 128, 155, 165, 178, 180, 210, 231, 241, 246, 258, 261 \\   & \\
$W3$   &   13, 14, 17, 41, 52, 54, 94, 101, 160, 180, 185, 228, 246, 249 \\ & \\
$W4$   &   13, 14, 41, 52, 64, 94, 128, 135, 170, 180, 189, 201, 204, 206, 210, 241, 242, 252 \\& \\
\hline
\end{tabular}
\caption{ \small List of multipoles with $P < 0.05$ for power entropy for
the 5-year WMAP-DA maps.}
\label{tab:tab1_5yr}
\end{center}
\end{table}

Fig.~\ref{pow_entropy} illustrates the entropy distributions leading to these $P$-values. A
contour for the $95\%$ confidence level is shown in gray. The 90\% and
50\%   confidence level contours are also shown as curves. The relatively large
spread of the distribution towards the small-$l$ region is kinematic, akin to cosmic variance.
The statistically anisotropic multipoles shown by red points are the same as those shown in Table~\ref{tab:tab1}.

\subsection{Significance: Power Entropy Statistics}

We now assess the significance of the numerous small $P$-values observed for the power entropy.

Table~\ref{tab:tab1} (\ref{tab:tab1_5yr})
shows 29 (24), 22 (20), 20 (19), 13 (16), 17 (20), 16 (19), 15 (14) and 15
(18)
power entropies with $P-value$ $\le 0.05$ for the 3-year (5-year)
$Q1$, $Q2$, $V1$, $V2$, $W1$, $W2$,
$W3$ and $W4$
maps respectively. The threshold values (upper bounds of $P$-values)
for
these  power entropies estimated using the individual maps are given by
$\mathcal P$ = 0.048 (0.047), 0.0467 (0.049), 0.049 (0.049), 0.0412 (0.048),
0.0438 (0.049), 0.0483 (0.047), 0.0472 (0.047), 0.0473 (0.049).
The total number of independent trials for $2   \leq l\leq 300$ is
$n=299$.   From the binomial distribution the {\it cumulative} probabilities of
obtaining $P_{bin}(k \geq k_{data}, P_{data}, \, 299)  $ are 
shown in Table \ref{tab:NetP}
for the eight maps from $Q1$ to $W4$ for the 3-year and 5-year data.

Clear violation of statistical isotropy is observed for $Q1$ and $Q2$ maps for both
the 3 and 5-year data.
which all have $P<0.05$. We noticed in our study that the $Q1$ and $Q2$ $P$-values are correlated over all $l$, so we cannot consider them independent.  Nevertheless the cumulative probability of $ 3 \times 10^{-4}$ for the $Q1$ band is far below anything expected from an isotropic ensemble.

\begin{figure}[htb]
\begin{center}
\includegraphics[width=4in]{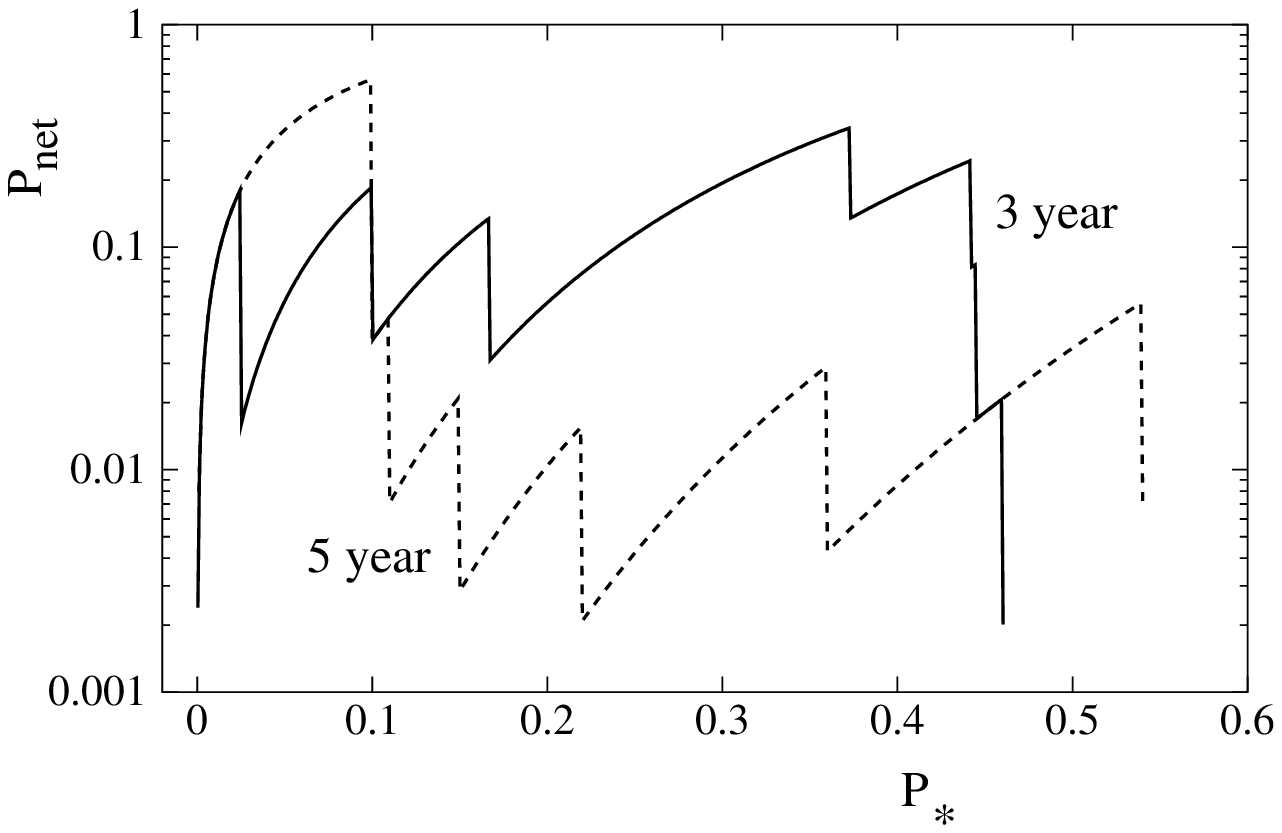}
\caption{The net probability $P_{\rm net}$ over all the bands selecting power-entropy $P_{band}<P_{*}$ for the three (solid line) and five (dashed line)
year WMAP data.}
\label{fig:PowPvalAssess}
\end{center}
\end{figure}

If one
assumes each probability is independent - which is certainly an idealization - the binomial probability for $Q1$ and $Q2$ for the 3 year data
to have such small probabilities is about $1.6 \times 10^{-2}$.
Fig. \ref{fig:PowPvalAssess} shows the probability of these outcomes over all bands as the ``pass-value'' $ P_{band}<P_{*}$ is adjusted for both the 3 and 5 year data. The small
$P_{\rm net}$ values show violation of isotropy. The entire data over all bands
shows violation of isotropy with a binomial
probability of $2.0\times 10^{-3}$
and $7.2\times 10^{-3}$ for the 3 and 5 year data respectively.

\begin{figure}[htb]
\centering
\includegraphics[scale=0.8,angle=0]{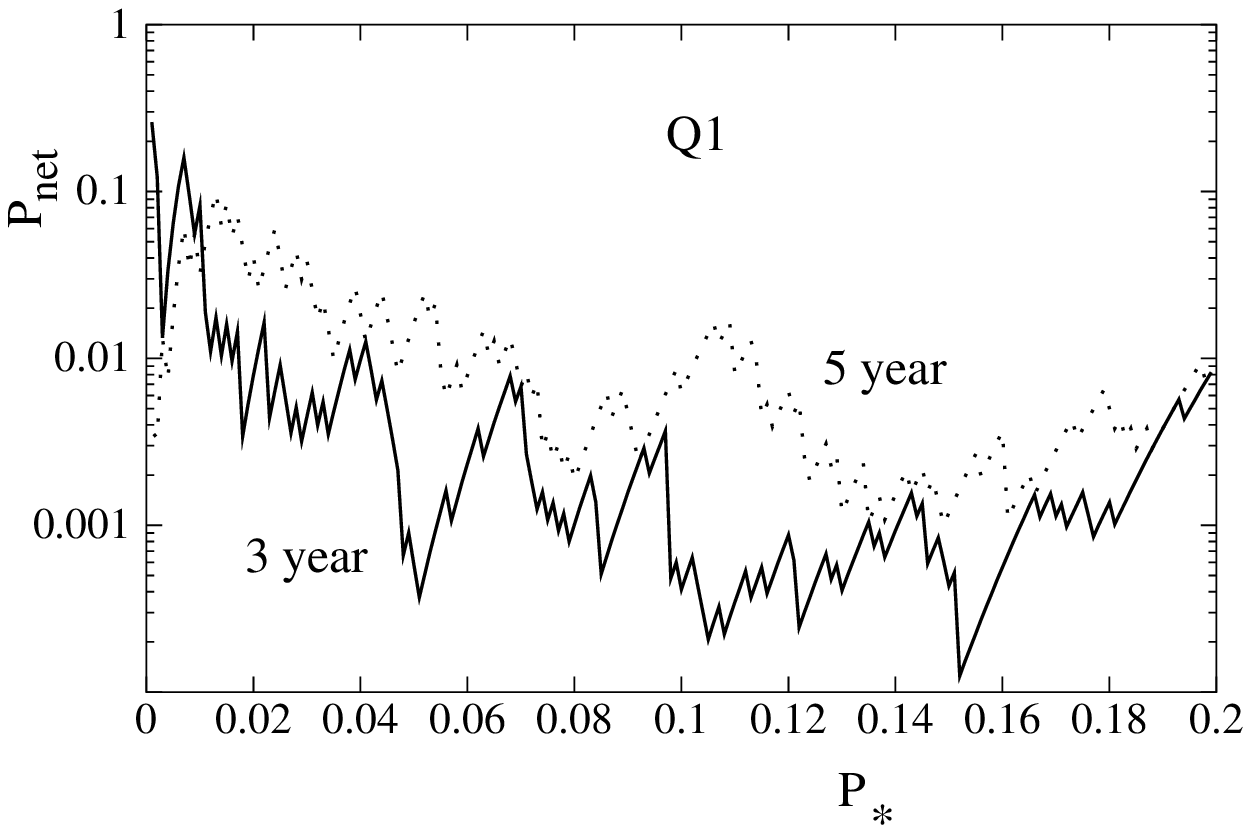}
\includegraphics[scale=0.8,angle=0]{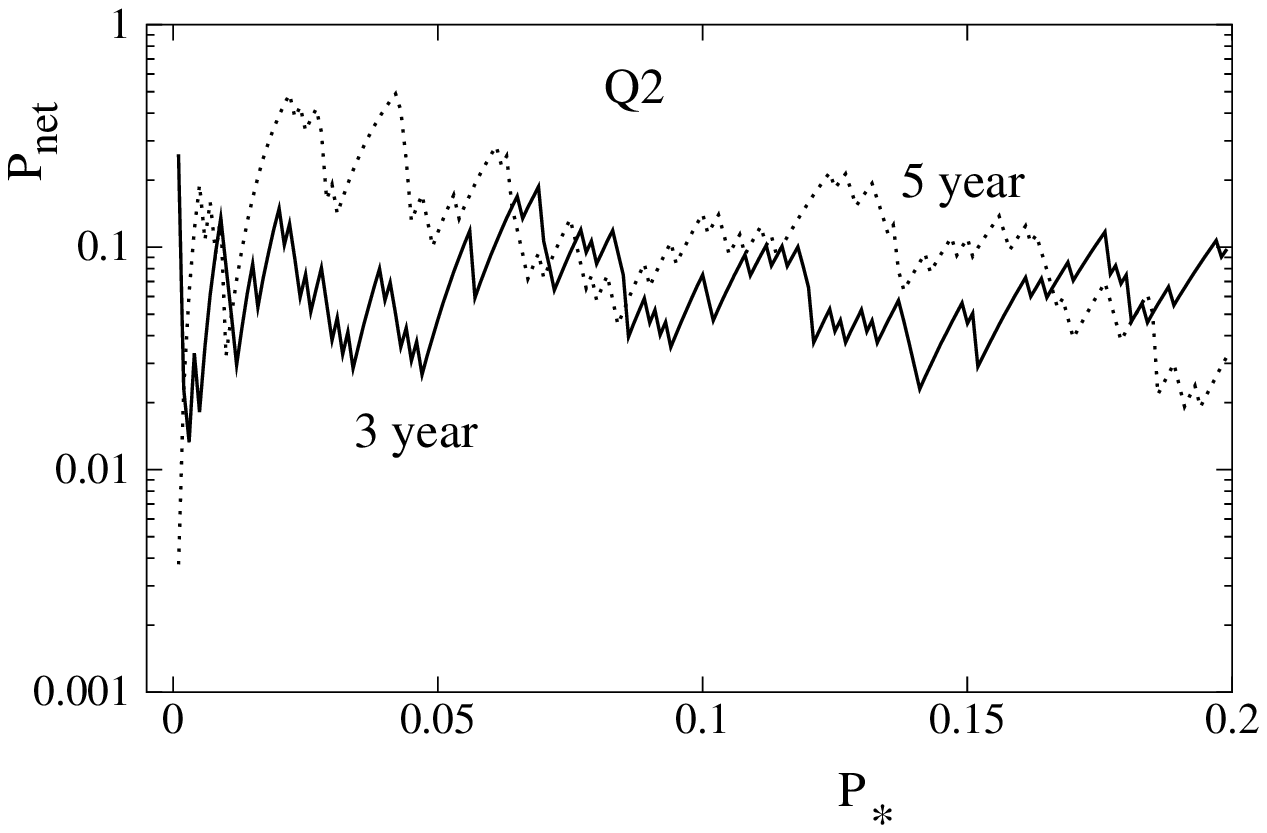}
\caption{The net cumulative probability $P_{\rm net}$ the DAs $Q1$ and 
$Q2$ selecting power-entropy $P_{band}<P_{*}$ for the three (solid line) and five (dotted line)
year WMAP data.   }
\label{fig:CumBinomiQ}
\end{figure}

\begin{figure}[htb]
\centering
\includegraphics[scale=1.0,angle=0]{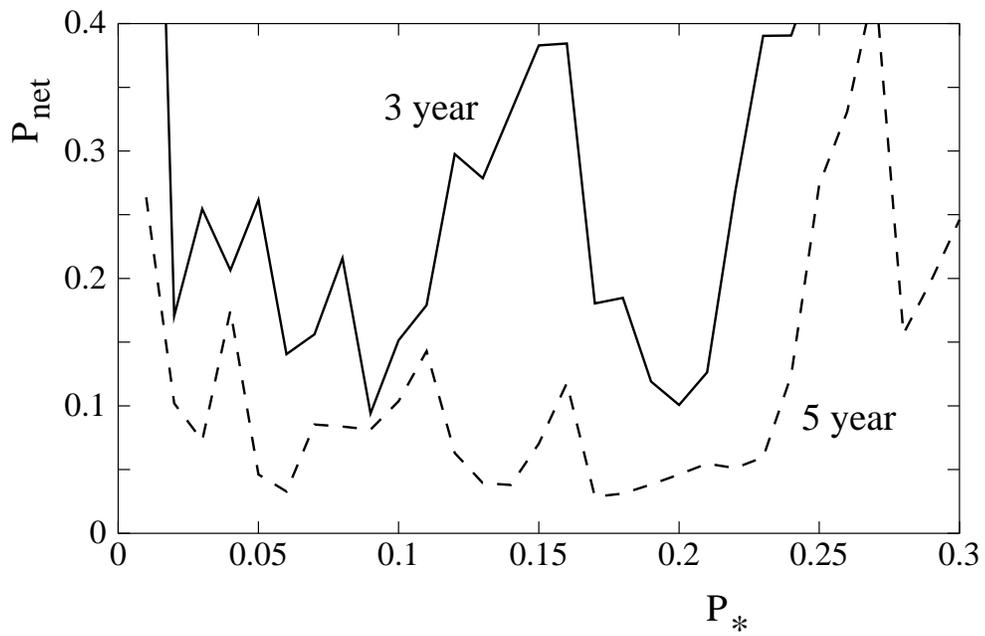}
\caption{The net cumulative probability $P_{\rm net}$ the DAs $V1$, $V2$, $W1$,
$W2$, $W3$ and $W4$ selecting power-entropy $P_{band}<P_{*}$ for the 
three (solid line) and five (dotted line)
year WMAP data.   }
\label{fig:CumBinomiVW}
\end{figure}

Since the 5 \% $P$-val cut is somewhat arbitrary, Fig. \ref{fig:CumBinomiQ} shows the cumulative probability of these outcomes over the $Q1$ and $Q2$ DAs
as the ``pass-value'' $ P_{band}<P_{*}$ is adjusted for both the 3 and 5-year data. The small
$P_{\rm net}$ values show violation of isotropy. The cumulative probability 
for the remaining six DAs is shown in Fig. \ref{fig:CumBinomiVW}. Here we 
notice that the 3-year data does not show a significant violation of isotropy.
However the signal of anisotropy is stronger in the 5 year data.
The trend in this figure suggests that we may expect a much stronger 
signal of anisotropy in $V$ and $W$ bands as more data is accumulated.

\begin{table}
\scriptsize
\begin{center}
\begin{tabular}{ccccc}
\hline
\hline
Band & $Q1$ & $Q2$ & $V1$ & $V2$ \\
Significance (3 year) & $ 3 \times 10^{-04} $ & $2.5 \times
10^{-02} $ & $ 0.10  $ & $ 0.46 $ \\
Significance (5 year) & $ 8.2 \times 10^{-03} $ & $0.10$ & $0.15$ & $0.36$ \\
\hline
Band & $W1$ & $W2$ & $W3$ & $W4$ \\
 Significance (3 year) & $ 0.17 $ & $ 0.37$ & $0.44 $ & $ 0.44 $\\
 Significance (5 year) & $ 0.10 $ & $ 0.11$ & $0.54 $ & $ 0.22 $\\
 \hline
\end{tabular}
\caption{\small Net significance of observing $P \leq 0.05$-values shown in Table \ref{tab:tab1} (3 year) and Table \ref{tab:tab1_5yr} (5 year).}
\label{tab:NetP}
\end{center}
\end{table}

\section{Alignment with the Quadrupole }
\label{sec:allign_l2}

Many authors (de Oliveira-Costa {\it et al} 2004, Ralston and Jain 2004,
Schwarz {\it et al} 2004) have observed a strong alignment between the 
CMB quadrupole and the octopole. The power of both quadrupole and 
octopole appears to
approximately lie in a plane. The perpendicular to the plane points roughly
in the direction of the Virgo supercluster for both these multipoles. It
has also been noted that these axes align closely with the CMB dipole,
as well as with independent cosmological observations.
Statistically significant alignment of several independent axes violates
the hypothesis of statistical isotropy. As reported earlier,
the WMAP-ILC map shows statistically significant signals of
alignment with the quadrupole axis in the
low $l$ multipole range $l\le 50$.

In our formalism one may construct
an unbiased measure of alignment between multipoles by comparing the
principal eigenvectors of the power tensor. In isotropic data these eigenvectors would point in random directions. The probability for isotropically distributed axes $\hat n$ and $\hat n'$ to align within $\theta$ is given
by \ba P(\cos \theta)= (1-\cos\theta) \ea
where
$\cos\theta =|\hat n\cdot \hat n' |$.

\begin{table}
\scriptsize
\centering
\begin{tabular}{cccccccc}
\hline
\hline
$Q1$       &   $Q2$           &   $V1$       &     $V2$   &   $W1$       &       $W2$     &       $W3$       &         $W4$     \\
\hline
\hline
  \\
2     &       2         &       2       &       2       &     2         &     2       &       2           &         2 \\
28   &       28       &       40     &       28     &     28       &     10     &       28         &       40 \\
61   &       61       &       50     &       61     &     61       &     28     &       61         &       61 \\
63   &       88       &       61     &       63     &     63       &     61     &       62         &       63 \\
75   &       101     &       66     &       66     &     81       &     63     &       63         &       88 \\
88   &       145     &       75     &       75     &     88       &     75     &       66         &       102\\
105 &       172     &       88     &       81     &     101     &     88     &       75         &       129\\
129 &       176     &       174   &       88     &     133     &     110   &       88         &       133\\
134 &       187     &       198   &       129   &     172     &     129   &     129         &       139\\
140 &       212     &       207   &       144   &     174     &     182   &     133         &       140\\
144 &                   &       226   &       172   &     176     &     197   &     177         &       243\\
145 &                   &       270   &       174   &     182     &     235   &     179         &       272\\
172 &                   &       278   &       182   &     267     &     267   &     265         &             \\
182 &                   &       289   &       187   &     279     &     270   &                     &             \\
        &                   &       293   &       207   &                 &               &                     &             \\
        &                   &                 &       243   &                 &               &                     &             \\
        &                   &                 &       267   &                 &               &                     &             \\
        &                   &                 &       279   &                 &               &                     &             \\
        &                   &                 &       293   &                 &               &                     &             \\
\hline
\end{tabular}
\caption{ \small List of multipoles with $P< 0.05$ for alignment with the
quadrupole for 3-year WMAP data for all the maps for
the multipole range $2 \le l \le 300$.}
\label{tab:tab2}
\end{table}

\begin{table}
\scriptsize
\centering
\begin{tabular}{cccccccc}
\hline
\hline
$Q1$       &   $Q2$           &   $V1$       &     $V2$   &   $W1$       &       $W2$     &       $W3$       &         $W4$     \\
\hline
\hline
  \\
2     &       2         &       2       &       2       &     2         &     2       &     2           &         2 \\
40   &       40       &       40     &       28     &     28       &     28     &     40         &         3 \\
42   &       42       &       42     &       40     &     40       &     40     &     50         &       40 \\
61   &       61       &       50     &       50     &     61       &     50     &     61         &       42 \\
75   &       88       &       61     &       61     &     63       &     61     &     75         &       61 \\
81   &     101       &       75     &       63     &     75       &     63     &     88         &       63 \\
88   &     134       &       88     &       66     &     81       &     75     &     133       &       88 \\
101 &     172       &     101     &       75     &     88       &     88     &     226       &       101 \\
105 &     176       &     129     &       88     &     101     &   133     &     236       &       129 \\
129 &     187       &     140     &       101   &     129     &   179     &     243       &       133 \\
134 &     195       &     174     &       129   &     133     &   182     &     265       &       139 \\
182 &     238       &     182     &       172   &     172     &   207     &     270       &       172 \\
      &                   &     207     &       174   &     174     &   267     &                   &       176 \\
      &                   &     279     &       182   &     178     &   270     &                   &       177 \\
      &                   &     300     &       187   &     182     &   171     &                   &       197 \\
      &                   &                 &       279   &     187     &               &                   &       243 \\
      &                   &                 &       293   &     234     &               &                   &       266 \\
      &                   &                 &                 &     267     &               &                   &       272 \\
      &                   &                 &                 &     278     &               &                   &                 \\
      &                   &                 &                 &     279     &               &                   &                 \\

\hline
\end{tabular}
\caption{ \small List of multipoles with $P< 0.05$ for alignment with the
quadrupole for 5-year WMAP data  over
the multipole range $2 \le l \le 300$.}
\label{tab:tab2_5yr}
\end{table}

\subsection{Significance of Axial Alignments}

Table~\ref{tab:tab2} (\ref{tab:tab2_5yr})
lists the multipoles with $P(\cos \theta) < 0.05$ for
alignment with the quadrupole for 3-year (5-year) WMAP  maps.
There are 13 (12), 9 (12), 14 (15), 18 (17), 13 (20), 13 (15), 12 (12) and
11 (18) axes which show alignment with the quadrupole moments for the
$Q1$, $Q2$, $V1$, $V2$, $W1$, $W2$, $W3$ and $W4$ maps respectively for 3-year (5-year)
data. The threshold values (upper bound of the $P$ values) are given by $\mathcal P$ = 0.046 (0.041), 0.049
(0.047), 0.038 (0.046), 0.048 (0.05), 0.048 (0.049), 0.048 (0.049), 0.044 (0.05), 0.049 (0.049). The binomial probabilities for each band are
respectively 0.62 (0.57), 0.96 (0.74), 0.25 (0.39),
0.19 (0.31), 0.68 (0.091), 0.68 (0.50), 0.66 (0.82), 0.87 (0.22)
for the 3-year (5-year) data.
Including the effects of the search over $2<l\le300$, the  set of multipole
axes examined shows no statistically significant signal of alignment.
We point out, however,
that the overall probabilities have a tendency to decrease as we go from three
to five year data.

There are several differences between the data
set used in the previous study and the one used for the present analysis.
The previous study used the ILC map, which is ideal for low $l$ multipoles.
This is because the ILC map has lower foregrounds and the entire map can be 
used. The
template cleaned maps are best suited for large $l$ multipoles and
require a mask to remove the contamination due to galactic and point source
emissions. In addition, the high $l$ data also contains very large
detector noise contamination, tending to decrease signal-to-noise.

%\begin{figure}[htb]
%\centering
%\includegraphics[width=4in]{poisson_quad.eps}
%\caption{ \small   The net probability $P_{\rm net}$ over all the bands for
%quadrupole alignment with $P_{band}<P_{*}$ for 3-year (solid line)
%and 5-year (dashed line) WMAP data. The ramps show the effects of removing
%small probabilities.}
%\label{fig:AlignPvalAssess}
%\end{figure}

\section{Alignment Entropy}

\begin{figure}
\centering
\includegraphics[scale=0.55,angle=-90]{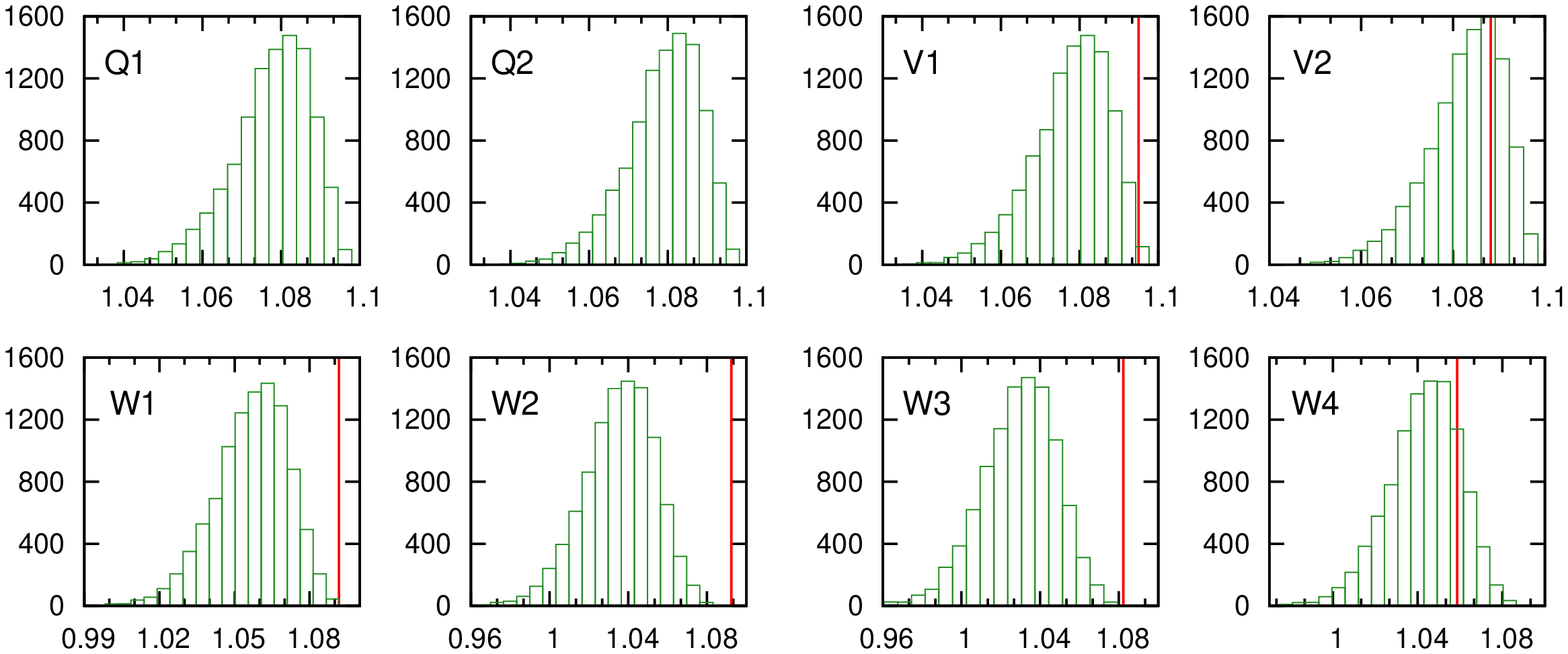}
\caption{ \small The distribution of the alignment entropy for the statistically isotropic $ CMB$ plus appropriate detector noise maps for the range $150 \le l\le 300$ for the WMAP 3-year data. The alignment entropy measures for different maps are also shown.}
\label{fig:X_entropy}
\end{figure}

\begin{figure}
\centering
\includegraphics[scale=0.55,angle=-90]{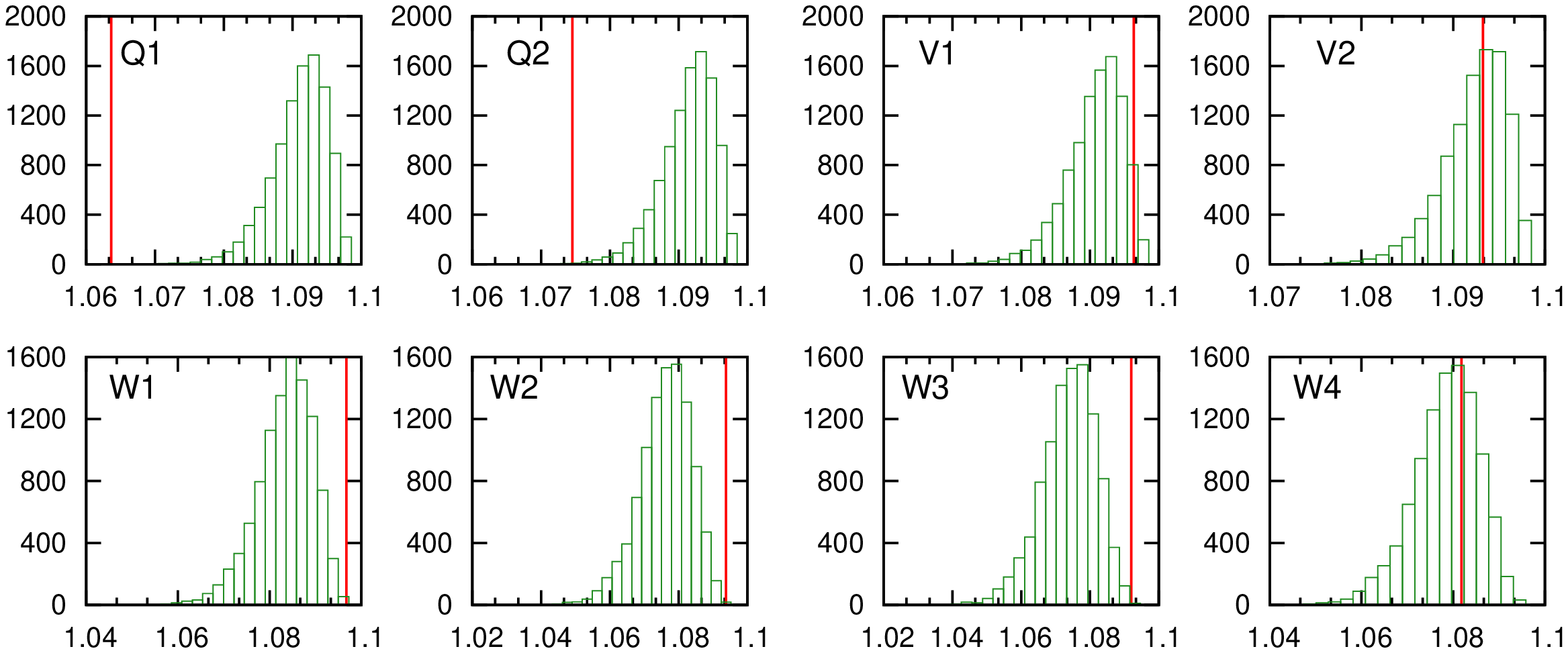}
\caption{ \small The distribution of the alignment entropy for the statistically isotropic $ CMB$ plus appropriate detector noise maps for the range $2 \le l \le 300$ for the WMAP 3-year data. The alignment entropy measures for different maps are also shown .}
\label{fig:X_entropy_2_300}
\end{figure}

We next consider the alignment entropy $S_{X}$ over  the entire range of
multipoles  $2\le l\le 300$, and a few selected subsets,
$150\le l\le 300$ and $250\le l\le 300$.
Figs. \ref{fig:X_entropy} and \ref{fig:X_entropy_2_300} show null
distributions of $S_{X}$ for the range $150\le l\le 300$ and $2\le l\le 300$. 
These distributions are
generated by simulated CMB data along with detector noise, appropriate
for a particular map.
The distributions of $S_{X}$ for the two cases are nearly identical. These distributions are similar to the power entropy distributions, consisting of
sharp suppression of small $S_{X}$ below a peak near the maximum.    The $S_{X}$ distributions for small $l$ show a long tail.
Figs. \ref{fig:X_entropy} and \ref{fig:X_entropy_2_300} also show the
value of $S_X$ obtained from the data for all cases except the maps $Q1$ and
$Q2$. For these two maps the value of $S_X$ lies outside the range shown
in the plots.

The values of $S_X$ for all the maps for the three year WMAP data are shown
in Table \ref{tab:tab3}. The probabilities of obtaining these values
from a random isotropic sample are also shown. These are computed by
using 10,000 randomly generated samples of isotropic $CMB$ maps including
detector noise.
The statistics are interesting. In all three sets  the
$Q$ band shows a very significant signal of violation of statistical
isotropy. The probability that the entropy obtained for $Q1$ map arises
by a random fluctuation is less than $0.01$ \% for all  three range
of multipoles considered. The map $Q2$ also shows very low probability values.

The preferred axes of alignment over the  different
ranges of multipoles are given in Table \ref{tab:tab4}.
We find that the axes do not point towards any familiar direction. The axes do
not point towards Virgo and hence are not aligned with the quadrupole. They 
tend to lie within about $30^o$ from the galactic plane at the galactic
longitude ranging between $90^o$ to about $100^o$. 
We next determine the mean axis in a simulated $Q1$
map in the range $2\le l\le 300$. Foregrounds are added
to this map by using the
publicly available Planck Sky Model (PSM)~\footnote{We acknowledge the use of version 1.1 of the Planck reference sky model, prepared by the
members of Working Group 2 and available at www.planck.fr/heading79.html.} as reference templates. We add foregrounds at the level of $1\%$, $2\%$, ..., $10\%$
of the total contamination and determine the mean vector for each map.
The mean vector is determined after applying the Kp2 mask and filling 
the masked region with randomly generated data, exactly as done for the real
data set.
As expected at low foreground level the mean axis fluctuates considerably
for different realizations of the randomly generated maps. However at 
foreground levels of 5 \% or higher, the mean axes stabilize. They also 
do not show much change with the increase in the level of
contamination. The axes are found to lie between $b=25^o-28^o$, $l=150^o-167^o$
for foreground levels of 5 \% or higher of their total values.

We compare the axes obtained using randomly generated maps
with the axes given in Table \ref{tab:tab4}. We
find that the galactic latitude matches well with that obtained from the
real data. However the longitude is off by almost $60^o-70^o$. Hence
it is not possible to assign the alignment we find to contamination due to 
known foregrounds. 
The randomly generated axes depend to some extent on the range of 
multipoles studied. For the multipole range $250\le l\le 300$, the 
mean axis is found to be roughly $b=6^o$, $l=125^o$. This is a little
closer to corresponding value in this range in Table \ref{tab:tab4}. 
We notice, however, that dependence of the axis on the choice of multipole range
is much stronger in the randomly generated data in comparison
to that found in Table \ref{tab:tab4}. This again shows that we cannot
attribute the anisotropy in $Q$ band to known foregrounds. 
It is possible that the anisotropy arises due to an unknown foreground
source or from a
combination of foregrounds and other effects.

The $V$ and $W$ bands
reveal an unexpected number of cases with very
{\it large} alignment entropy, corresponding to unusually {\it perfect}
isotropy. We find several cases in the $W$ band where the
alignment entropy is so large that the probability to obtain this from
a random sample exceeds $99.99$ \%.

\begin{figure}
\centering
\includegraphics[width=4in]{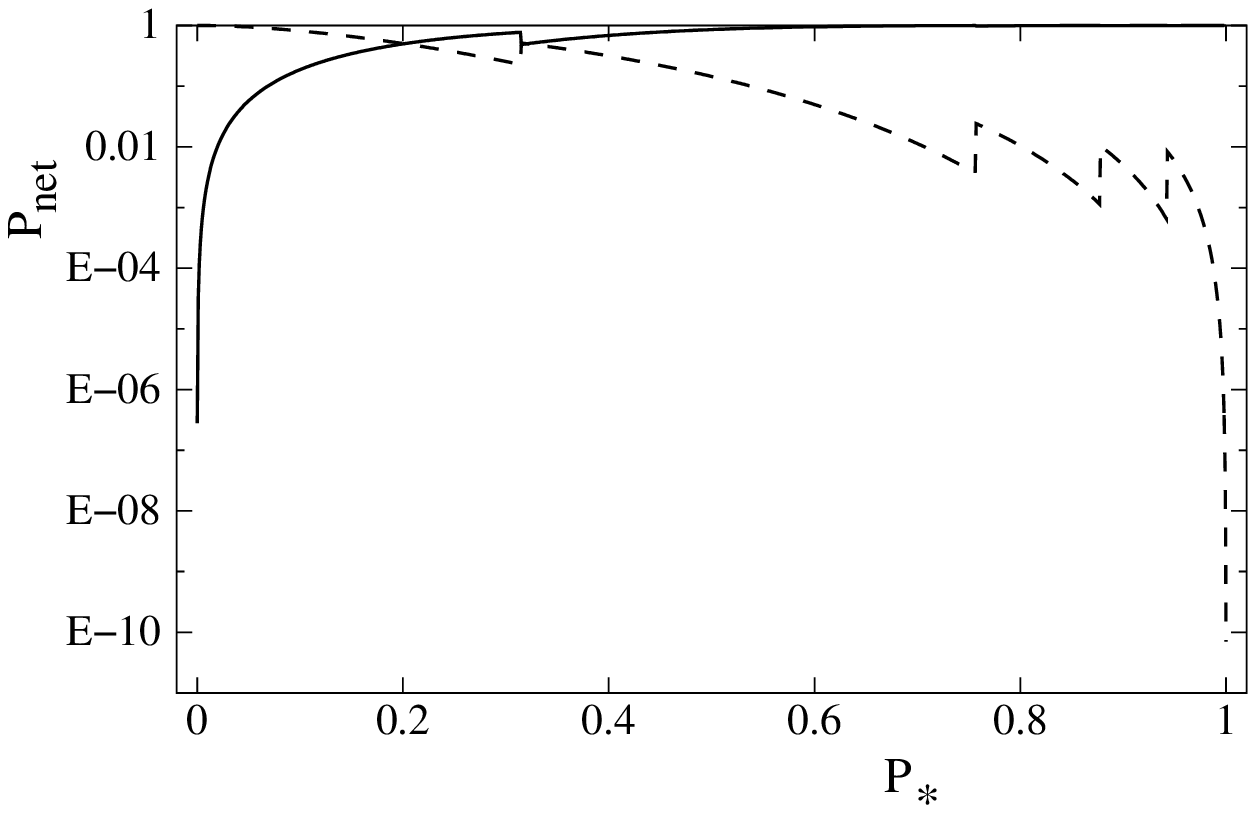}
\caption{ \small The net probability $P_{\rm net}$ across bands to find excessive anisotropy (solid
line) and isotropy (dashed line) with the alignment entropy for the WMAP
five year data. Multipole range $150\le l\le 300$. }
\label{fig:AlignPValRows}
\end{figure}

\begin{table}
\scriptsize
\begin{tabular}{ccccccccc}
% after \\ : \hline or \cline{col1-col2} \cline{col3-col4} ...
\hline
\hline
 $Q1$   & $Q2$   & $V1$   & $V2$   & $W1$   & $W2$ & $W3$   & $W4$   \cr
\hline
$S_X(150, 300 )$ &   0.98522 & 1.02024 & 1.09499 & 1.08822 & 1.09174 & 1.09234  & 1.08222 & 1.05858 \cr
$P(\%) $ & $<$0.01 & $<$0.01 & 99.25 & 72.08 & 99.98 & $>$99.99 & $>$99.99 & 82.10 \cr
\hline
$S_X(250, 300 )$ & 0.79763 & 0.92503 & 1.077021 & 1.08415 & 1.055178 & 1.07635 & 0.98258 & 1.019228 \cr
$P(\%)$ & $<$0.01 & 0.36 & 94.72 & 94.89 & 97.36 & 99.9 & 86.56 & 92.03 \cr
\hline
$S_X(2, 300 )$ &   1.0636 & 1.0745 & 1.0964 & 1.0932 & 1.0967 & 1.0937 & 1.0920 & 1.0818 \cr
$P(\%) $ & $<$0.01 & 0.15 & 95.86 & 53.94 & 99.94 & 99.74 & 99.91 & 65.52 \cr
\hline
\end{tabular}
\caption{ \small Alignment entropy $S_{X}$ and corresponding P values (in \%)
for WMAP 3-year maps
over the three multipole ranges, $2\le l\le 300$,
$150\le l\le 300$ and $250\le l\le 300$.
}
\label{tab:tab3}
\end{table}

\begin{table}
  \centering
  \begin{tabular}{cccc}
\hline
\hline
$Q1$ band & $b$ (deg) & $l$ (deg) \\
\hline
\hline
$150 \le l \le 300$ & 27.8 & 97.8 \\
$250 \le l \le 300$ & 30.2 & 101.5 \\
$2 \le l \le 300$ & 24.7   & 92.9 \\
\hline
\hline
$Q2$ band & $b$ (deg)& $l$ (deg)\\
\hline
\hline
$150 \le l \le 300$ & 26.3 & 94.6 \\
$250 \le l \le 300$ & 28.1 & 99.2 \\
$2 \le l \le 300$ & 22.2 & 89.7 \\
\hline
\hline
\hline
\hline
\end{tabular}
\caption{The galactic latitude ($b$) and longitude ($l$) for the principal
axis for the specified range of multipole moments for WMAP 3-year
$Q1$ and $Q2$ bands}
\label{tab:tab4}
\end{table}

Similar results are seen for the five year WMAP data.
In Table \ref{tab:tab5} we show the alignment entropy $S_X$
and probabilities $P$ for all the maps in the three multipole ranges
considered. We again find that the $Q$ band shows a very striking
signal of anisotropy. The $W$ band, on the other hand, again shows an improbablly high
level of isotropy. The $V$ band does not  appear statistically unusual.
Table \ref{tab:tab6} shows the axes of alignment for the $Q$ band.
The axes are found to be consistent with that found in the three year data.

Fig. \ref{fig:AlignPValRows} shows the net probability across bands for $P<P_*$
or ``excessive anisotropy" as well as $P>P_{*}$, or ``excessive isotopy'' 
for the 5 year data.

\begin{table}
\scriptsize
  \begin{tabular}{ccccccccc}
%% after \\ : \hline or \cline{col1-col2} \cline{col3-col4} ...
\hline
\hline
& $Q1$   & $Q2$   & $V1$   & $V2$   & $W1$   & $W2$ & $W3$   & $W4$   \cr
\hline
$S_X(150, 300 )$ &   1.00795 & 1.02283 & 1.09116 & 1.08587 & 1.08633 & 1.09417 & 1.09522 & 1.08451 \cr
$P(\%) $ & $<$0.01 & $<$0.01 & 75.7 & 31.5 & 87.8 & $>$99.99 & $>$99.99 & 94.3 \cr
\hline
$S_X(250, 300 )$ & 0.85946 & 0.89287 & 1.08185 & 1.08579 & 1.08321 & 1.08854 & 1.05246 & 1.07737 \cr
$P(\%)$ & $<$0.01 & $<$0.01 & 86.0 & 85.9 & 98.8 & 99.9 & 96.9 & 99.4 \cr
\hline
$S_X(2, 300 )$ &   1.07100 & 1.07602 & 1.09462 & 1.09283 & 1.09362 & 1.09259 & 1.09694 & 1.08896 \cr
$P(\%) $ & $<$0.01 & $<$0.01 & 60.1 & 28.6 & 83.3 & 93.6 & 99.98 & 61.6 \cr
\hline
\end{tabular}
\caption{The alignment entropy and the corresponding P values (in \%)
for the WMAP 5-year DA maps. The results for all the three multipole ranges
considered in this paper are shown.
}
\label{tab:tab5}
\end{table}

\begin{table}
  \centering
  \begin{tabular}{ccc}
\hline
\hline
$Q1$ band & $b$ (deg)& $l$ (deg)\\
\hline
\hline
$150 \le l \le 300$ & 24.6 & 98.2 \\
$250 \le l \le 300$ & 27.3 & 103.2 \\
$2 \le l \le 300$ & 20.2 & 93.1 \\
\hline
\hline
$Q2$ band & $b$ (deg)& $l$ (deg)\\
\hline
\hline
$150 \le l \le 300$ & 29.6 & 94.7 \\
$250 \le l \le 300$ & 32.2 & 97.3 \\
$2 \le l \le 300$ & 24.0 & 88.9 \\
\hline
\hline
\end{tabular}
\caption{The galactic latitude ($b$) and longitude ($l$) for the principal
axis for the specified range of multipole moments for WMAP 5-year $Q1$ and $Q2$ bands}
\label{tab:tab6}
\end{table}

\subsection{Foreground contamination in $Q$ band}
One might naturally assume the anisotropy found in the $Q$ band  would be due to foreground contamination. The principal vectors for all
the multipole ranges considered here cannot be consistently attributed to known
foregrounds. 
Let us nevertheless assume that foregrounds give a
significant contribution to the $Q$ band anisotropy,
and seek the mean foreground power required to explain the observations.
We restrict this study to the multipole range
$150 \le l \le 300$.

To estimate residual foreground contamination
in the maps we use PSM as reference templates. 
We first generate a composite
foreground map corresponding to each map using synchrotron, dust and free-free maps obtained by PSM. We apply the $Kp2$ mask
to all the composite foreground maps also
in order to avoid strong contamination arising from the galactic region.
Finally we add a small fraction of the
composite foreground contamination arising from these masked templates to a randomly generated $CMB$ map, plus simulated detector noise appropriate 
to each maps. We finally compute the alignment entropy for each band.

The residual foreground contamination in regions not affected by the $Kp2$ 
mask was estimated from the fraction of the composite masked foreground template added to randomly generated $CMB$ maps. We obtain the full-sky estimates of the foreground contamination using the MASTER method (Hivon {\it et al} 2002) which
employs inversion of the mode-mode coupling matrix to convert the partial-sky power spectrum to full-sky estimates.

\begin{figure}
\centering
\includegraphics[scale=0.55,angle=-90]{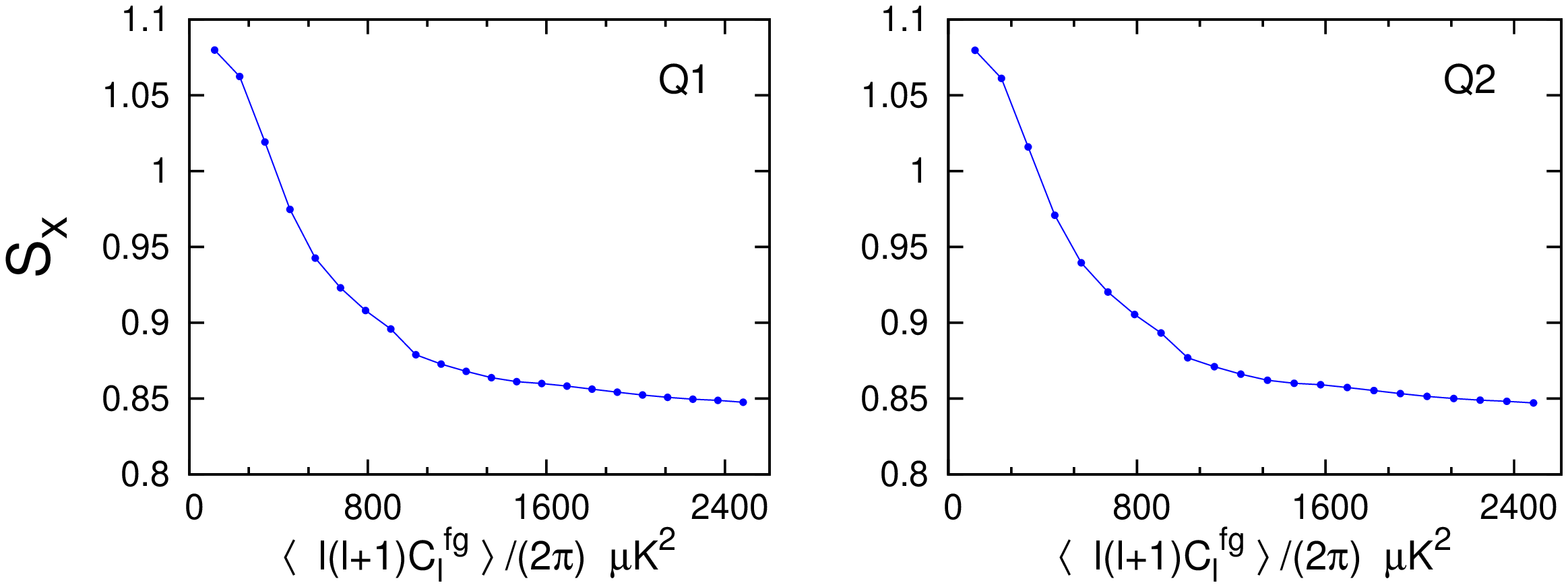}
\caption{ \small The alignment entropy, $S_X$,  for the bands $Q1$ and $Q2$
for the multipole range $150\le
l \le 300$ as a function of the average
foreground power (see text). }
\label{fg_power_vs_entropy}
\end{figure}

\begin{figure}
\centering
\includegraphics[scale=0.45,angle=0]{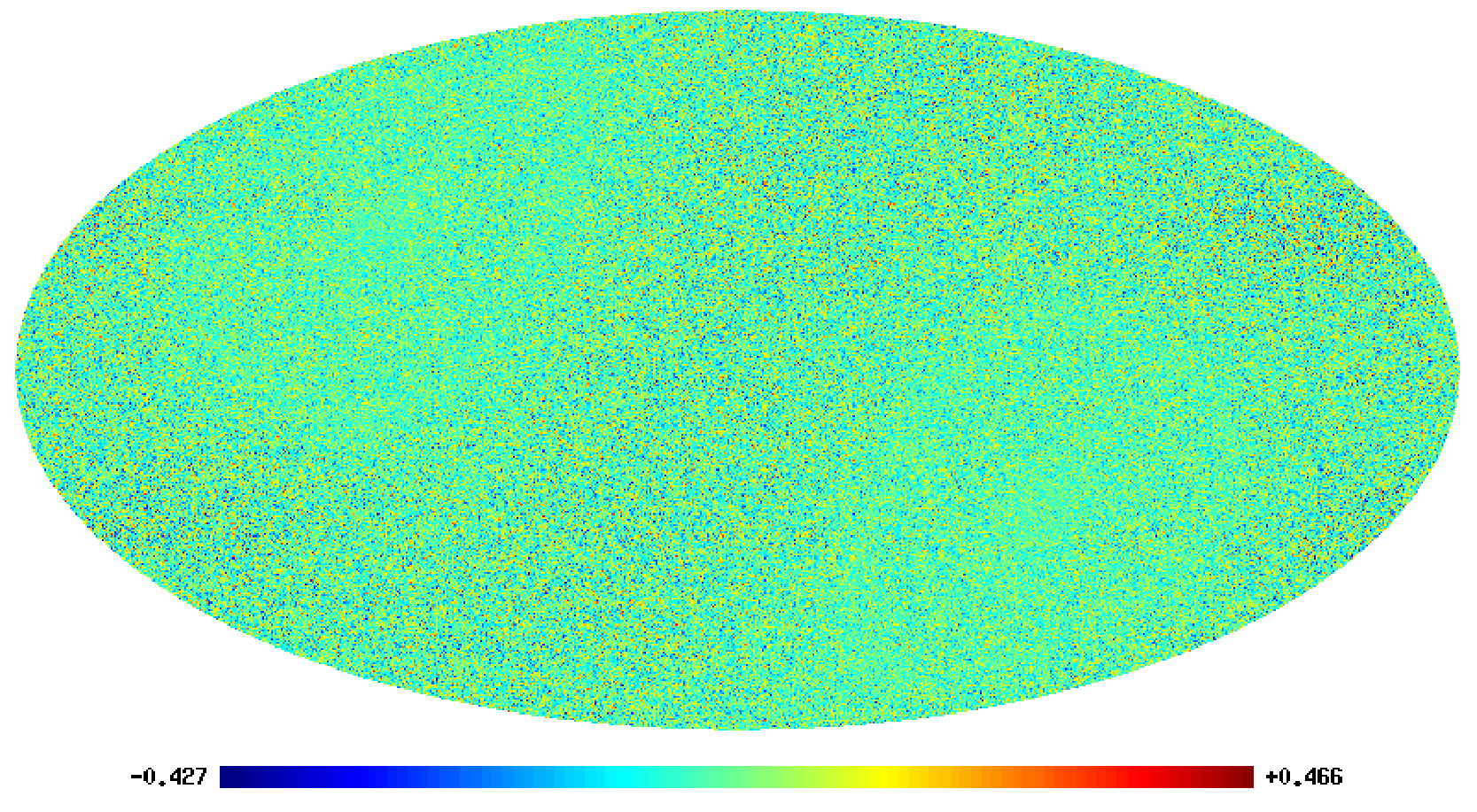}
\includegraphics[scale=0.45,angle=0]{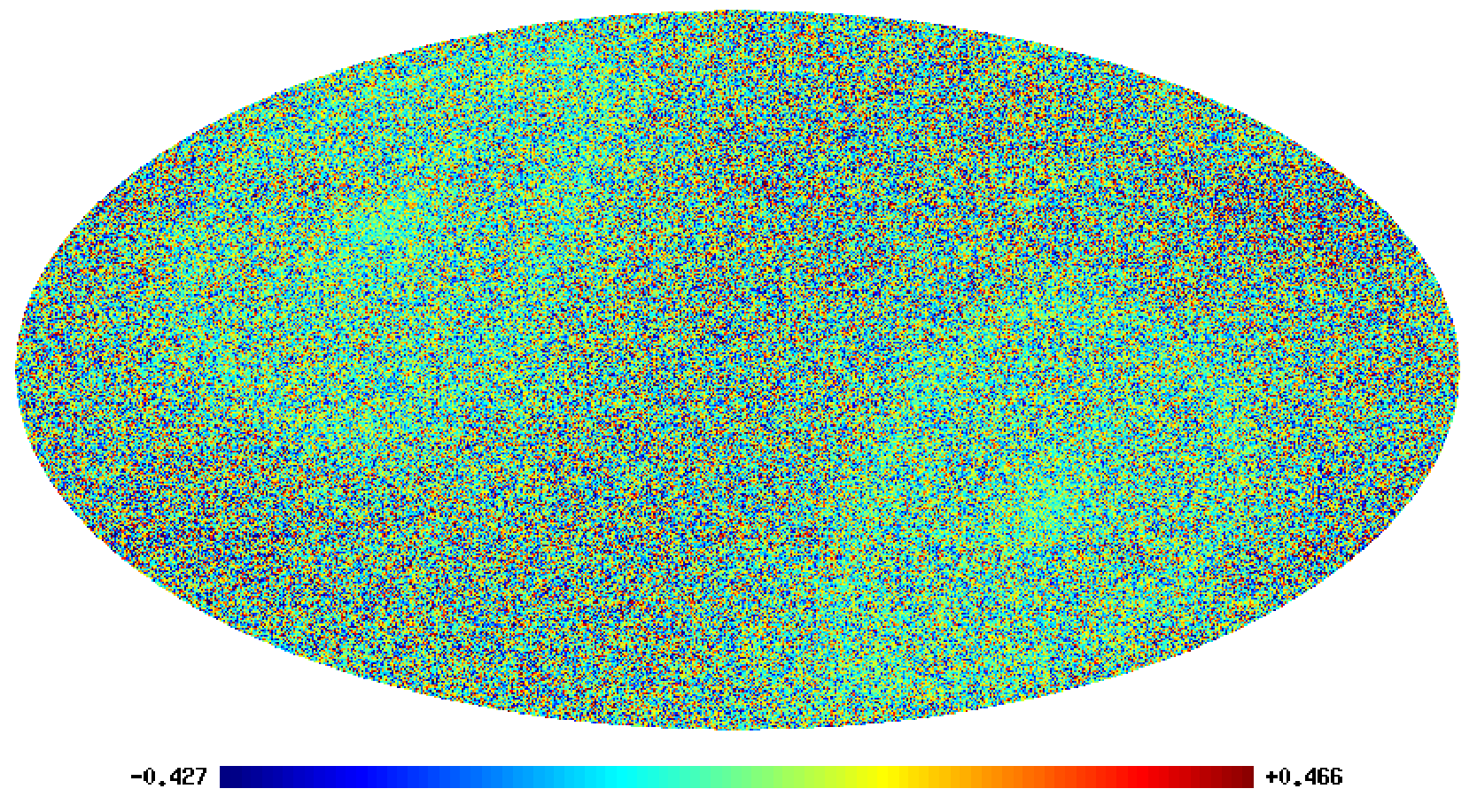}
\caption{ \small The generated noise maps for $Q1$ (upper) and $W4$ (lower)
bands.}
\label{noisemaps}
\end{figure}

\begin{table}
\scriptsize
  \begin{tabular}{lcc}
% after \\ : \hline or \cline{col1-col2} \cline{col3-col4} ...
\hline
\hline
maps & $Q1$   & $Q2$   \cr
\hline
Average         &                 &               \cr
Foreground   & 420.72 (3-year)   & 330.48 (3-year)   \cr
Power $(\mu K^2)$     & 417.33 (5-year)   & 375.73 (5-year) \cr
%\hline
%Average             &                 &               &             &             &             &             &               &               \cr
%Foreground       & 2521.27       &       633.91 &             81.48 &             62.01   &           123.14   &           83.81 &             220.24 &             179.01 \cr
%Power $(\mu K^2)(250 \le l \le 300) $       &                 &               &             &             &             &             &               &               \cr
\hline
\end{tabular}
\caption{ \small The average foreground residual power, $<l(l+1)C^{fg}_l>/(2\pi)$,
for the WMAP 3-year and 5-year $Q1$ and $Q2$ maps, which show significant signals of anisotropy with $P\le 0.01$ \% for the multipole range
$150 \le l \le 300$. The foreground power has been averaged over this range
of multipoles, as explained in text.}
\label{tab4}
\end{table}

In Fig.~\ref{fg_power_vs_entropy} we show the alignment entropy as a function of the
average value of the full-sky estimates of the residual foreground contamination for each band for the multipole range $150 \le l \le 300$.
We estimate the average foreground power for the range of multipole
moment $l_{min} \le l \le l_{max}$ as,
\begin{equation}
<l(l+1)C_l^{fg}> = {1\over (l_{max}-l_{min}+1)} 
\sum_{l=l_{min}}^{l_{max}} l(l+1)C^{fg}_l\, ,
\end{equation}
where $C_l^{fg}$ is the foreground power spectrum at $l$. 
For a given value of the entropy obtained from the data this figure   gives  the   average level of residual foreground contamination
in the range of multipoles under consideration.  The $Q1$ and $Q2$ maps indicate a strong level of foreground contamination for the multipole range $150 \le l \le 300$. Table~\ref{tab4}
shows the estimated residual foreground contamination quantitatively.

\subsection{Isotropy in $V$ and $W$ bands}
The very striking result seen in Table~\ref{tab:tab3} is the unusually high 
P-values for many of the  multipoles in the $V$ and $W$ 
bands for the three year
WMAP data. This anomaly is 
also supported by the WMAP five year data for the $W$ band. This is very 
unexpected and shows a statistically  unusual high level of isotropy. 
We are unable to identify the cause of this 
anomaly. One possibility is the neglect of noise correlations in our 
analysis. The anomaly is ameliorated if we artificially lower the detector
noise level in the simulated random maps.
The $\sigma_0$ values used for generating the noise maps for
the bands $Q1, \, Q2, \, V1, \, V2, \, W1, \, W2, \, W3, \, W4$ 
 are
 2.245, 2.135, 3.304, 2.946, 5.883, 6.532, 6.885, 6.744,  respectively.
Fig. \ref{noisemaps}   shows the generated noise maps for the 
bands $Q1$ and $W2$. The $W2$ map over the range $150\le l\le 300$ shows a $P$-value of 100\%. To explore this, we studied
how the P-value changes using a smaller value of 
$\sigma_0$ 
Reducing $\sigma_0$ by two units 
to $4.532$, the P-value decreases to a
more reasonable value of 92\%. However we find such a large change in the value
of $\sigma_0$  unacceptable. The problem of statistically unlikely isotropy is not solved 
in the present paper.

\section{Conclusions}
\label{conclu}
The possible violation of isotropy in CMB has been a subject of intense 
research after the publication of WMAP data. The possible alignment of axes
corresponding to several diverse data sets in the direction of the Virgo
cluster makes this extremely interesting. Despite several proposals the
origin of this effect is so far unknown. 

We have developed a general method
to test for statistical isotropy in the CMB data. The method assigns three 
orthogonal eigenvectors and the corresponding eigenvalues for each $l$ 
multipole. The dispersion in the eigenvalues is quantified by defining 
the concept of power entropy and provides a measure of the violation of
statistical isotropy. The principal eigenvector, i.e. 
the eigenvector corresponding
to maximum eigenvalue, can also be compared across different multipoles.
This yields another measure of violation of isotropy. We also
define the concept of alignment entropy which tests for dispersion in the
principal eigenvectors across a range of $l$ values.
We apply these techniques to the foreground cleaned DA maps provided 
by the WMAP team for their 3-year and 5-year data. 

We find that 
some of the DA maps, particularly those corresponding to $Q$ band, show
signal of significant violation of statistical isotropy. 
We are unable to attribute this violation to known foreground 
contamination. Assuming that the signal arises dominantly due to foregrounds,
we obtain an estimate of the residual foreground contamination in these
maps. We also find a significant signal of anisotropy if we combine 
the results obtained from all the DAs. The $V$ and $W$ band do not by 
themselves yield a significant signal of anisotropy. However the violation
of isotropy in these DAs is much stronger in the 5 year data in comparison 
to the 3 year data. This suggests that the signal of anisotropy in these 
data sets may be masked by the presence of large detector noise and may
become much more significant as we accumulate more data.

We do not find a signal of significant alignment with the quadrupole
in the present data. In an earlier paper (Samal {\it et al} 2008) we
did find a significant signal in the ILC map in the low multipole range.
In this range of multipoles the ILC map is most reliable.
This leads us to conclude that alignment with the quadrupole may be present
only at low multipoles. The presence of residual foregrounds and 
detector noise in individual DA maps, however, may hide a signal of alignment.
In our studies using alignment entropy we find a highly significant signal 
of anisotropy for the $Q$ band. This is consistent with the results we found
using power entropy. A conservative interpretation is that the $Q$ band 
anisotropy arises due to residual foregrounds. However we are unable to 
attribute the alignment in the $Q$ to known foregrounds.
 The principal axes, for all the 
multipole ranges considered, are consistent with one another and do not agree
well with those found by using simulated data with PSM foreground templates.
Our results indicate the existence of some unknown foreground contamination
or some other effect.

In the $W$ band we find an improbable level of isotropy in the data. This is 
quite unexpected. We considered whether there might be  due to incorrect
assumptions in our random simulations. Yet the assumptions we make are standard.
Excess isotropy appears to be a serious problem. This has implications  beyond
the issues addressed here.  It would be interesting to test the common 
assumption that detector noise is inherently uncorrelated. The question is 
important since incorrect modelling of detector noise
may also lead to bias in the estimation of CMB power and the
cosmological parameters.

\section{Acknowledgments}
We acknowledge the use of Legacy Archive for
Microwave Background Data Analysis. Some of the results of this work
are derived using the publicly available HEALPIx package 
(G\'{o}rski {\it et al}, 2005). Pramoda K. Samal acknowledges CSIR, India for financial support under the research grant CSIR-SRF-9/92(340)/2004-EMR-I. John P.
Ralston is
supported in part under DOE Grant Number DE-FG02-04ER14308. A portion of the research described in this paper was carried out at the Jet Propulsion Laboratory,
California Institute of Technology, under a contract with the National Aeronautics and Space Administration.

\bigskip

\bf{\large  References}

\begin{itemize}

\item[] Abramo, L. R.,  Sodre Jr., L., Wuensche, C. A., 2006,
Phys. Rev. D74, 083515

\item[]  Armendariz-Picon, C., 2004, 
JCAP 0407, 007

\item[]  Armendariz-Picon, C., 2006, 
JCAP 0603, 002

\item[] Battye, R. A.,  Moss, A., 2006, Phys. Rev. D74, 04130

\item[] Berera, A., Buniy, R. V., Kephart, T. W., 2004, JCAP 0410, 016

\item[] Bernui, A., 2008, arXiv:0809.0934 

\item[] Bernui, A., Villela, T., Wuensche, C. A., Leonardi, R.,
 Ferreira, I., 2006, Astron. Astrophys. 454, 409

\item[] Bernui, A., Mota, B., Reboucas, M. J., Tavakol, R., 
2007, Astron. Astrophys. 464, 479

\item[]  Bietenholz, M. F.,   Kronberg, P. P., 1984,   ApJ,
 {287}, L1-L2 

\item[] Bielewicz,  P., G\'{o}rski,  K. M., Banday, A.J., 2004, 
MNRAS, 355, 1283

\item[] Bielewicz, P., Eriksen, H. K., Banday, A. J., G\'{o}rski, K. M.,
 Lilje,  P. B., 2005, ApJ 635, 750 

\item[]
 Birch, P., 1982,   Nature,  298, 451 

\item[] Boehmer, C. G., Mota, D. F., 2008, Phys. Lett. B663, 168

\item[] Buniy, R. V., Berera, A., Kephart, T. W., 2006, Phys. Rev. D73, 063529 

\item[]  Campanelli, L., Cea, P., Tedesco, L., 2007, arXiv:0706.3802

\item[] Cline, J. M.,  Crotty, P.,  Lesgourgues, J., 2003, JCAP 0309, 010,
astro-ph/0304558;  

\item[] Contaldi, C. R.,  Peloso, M,  Kofman, L.,  Linde, A., 2003, 
JCAP 0307, 002, astro-ph/0303636  

\item[] Copi, C. J., Huterer, D., Schwarz, D. J,  Starkman, G. D., 2006,
MNRAS 367, 79

\item[] Copi, C. J., Huterer, D., Schwarz, D. J,  Starkman, G. D., 2007,
  Phys.\ Rev.\  D75, 023507 
  
\item[] de Oliveira-Costa, A.,  Tegmark, M., Zaldarriaga, M., 
Hamilton, A., 2004, Phys. Rev. D 69, 063516,
astro-ph/0307282

\item[]  de Oliveira-Costa, A.,  Tegmark, M., 2006, 
Phys. Rev. D74, 023005

\item[]  Dennis, M. R., 2005, J. Phys. A 38, 1653, arXiv:math-ph/0410004

\item[] Dimopoulos, K., Lyth, D. H., Rodriguez, Y., 2008, arXiv:0809.1055 

\item[] Donoghue E. P.,  Donoghue, J. F., 2005, Phys. Rev. D71, 043002

\item[] Donoghue, J. F., Dutta, K. and Ross, A., 2007, arXiv:astro-ph/0703455

%\item[] Dunkley, J. {\it et al}, 2008, arXiv:0803.0586

\item[] Efstathiou, G., 2003, MNRAS  346, L26,
astro-ph/0306431

\item[]  Erickcek, A. L., Kamionkowski, M., Carroll, S. M., 2008, 
arXiv:0806.0377

\item[] Eriksen, H. K. {\it et al}, 2004, ApJ  605, 14

\item[] Eriksen, H. K. {\it et al}, 2007a,  ApJ  656,  641

\item[] Eriksen, H. K., Banday, A. J., G\'{o}rski, K. M., 
Hansen, F. K.,  Lilje, 
P. B., 2007b, ApJ,  660, L81

\item[] Freeman, P. E., Genovese, C.R., Miller, C.J., Nichol, R.C., 
Wasserman, L., 2006, ApJ, 638, 1

\item[] Gaztanaga, E., Wagg, J., Multamaki, T., Montana, A.,  Hughes, D. H., 2003, MNRAS 346, 47, astro-ph/0304178 

\item[]  Gordon, C., Hu,  W., Huterer, D.,  Crawford, T. 2005, 
Phys. Rev. D72, 103002

\item[] G\'{o}rski, K.M., Hivon, E., Banday, A.J., Wandelt, B.D., Hansen, F.K.,
 Reinecke, M., \& Bartelmann, M., 2005, ApJ, 622, 759

\item[] Gumrukcuoglu, A. E., Contaldi, C. R., Peloso, M., 2006, 
astro-ph/0608405

\item[] Gumrukcuoglu, A. E., Contaldi, C. R., Peloso, M., 2007, 
arXiv:0707.4179

\item[] Hajian, A., Souradeep,  T., Cornish,  N. J., 2004, 
ApJ 618, L63

\item[]  Hajian, A.,  Souradeep, T., 2006, Phys. Rev. D74, 123521

\item[]  Hansen, F. K., Banday, A. J., G\'{o}rski, K. M., 2004, MNRAS, 
354, 641

\item[] Helling, R. C., Schupp, P., Tesileanu T., 2007, Phys. Rev. D74,
063004

\item[] Hunt, P., Sarkar, S., 2004, Phys. Rev. D70, 103518

\item[] 
Hinshaw, G. {\it  et al.}, 2003, Astrophys. J., Suppl. Ser.  148, 63

\item[] Hinshaw, G. {\it et al}, 2007, ApJ Suppl 170, 288 

\item[]  Hivon, E., {\it et al}, 2002,  ApJ,  567, 2

\item[] Hutsem\'{e}kers, D., 1998 {\it A \& A},   332, 41 

\item[]  Hutsem\'{e}kers, D., Lamy, H., 2001,  {\it A \& A},  {367}, 381 

\item[] Hutsem\'{e}kers, D., Payez, A.,  Cabanac, R., Lamy, H., Sluse, D.,
 Borguet, B., Cudell,  J.-R., 2008, arXiv:0809.3088

\item[] Inoue, K. T., Silk, J., 2006, ApJ 648, 23 

\item[] Jain, P.,  Narain, G.,   Sarala, S., 2004, {\it MNRAS},  347, 394 

\item[] Jain, P.,  Panda, S., Sarala, S., 2002, Phys. Rev. D66, 085007 

\item[] Jain, P,  Ralston, J. P., 1999,  Mod. Phys. Lett. A14, 417 

\item[]  Jain, P.,  Sarala, S., 2006, J. Astrophysics and Astronomy,  27, 443 

\item[] Land, K.,  Magueijo, J., 2005, Phys. Rev. D72, 101302 

\item[] Land, K.,  Magueijo, J., 2006, MNRAS 367, 1714 

\item[] Land, K., Magueijo, J., 2007, MNRAS 378, 153 

\item[] Lew, B., 2008, arXiv:0808.2867 

\item[] Liu, H., Li, T-P, 2008, arXiv:0806.4493

\item[] Kahniashvili, T.,  Lavrelashvili, G., Ratra, B., 2008, arXiv:0807.4239 

\item[] Kanno, S., Kimura, M., Soda, J., Yokoyama, S., 2008, JCAP 0808, 034

\item[] Katz, G., Weeks,  J., 2004, Phys. Rev. D70, 063527

\item[] Kendall, D. G., Young, A. G., 1984,  MNRAS,  {207}, 637 

\item[] Kesden, M. H., Kamionkowski, M., Cooray, A., 2003,  
Phys. Rev. Lett. 91,  221302, astro-ph/0306597  

\item[] Koivisto, T., Mota, D. F., 2006, Phys. Rev. D73, 083502

\item[] Koivisto, T., Mota, D. F., 2007, arXiv:0707.0279

\item[] Magueijo, J.,  Sorkin, R. D., 2007, MNRAS Lett. 377, L39

\item[] Moffat, J. W., 2005, JCAP 0510, 012

\item[] 
Naselsky, P. D., Verkhodanov, O. V.,  Nielsen, M. T. B., 2007,
arXiv:0707.1484

\item[] Payez, A., Cudell, J. R., Hutsem\'{e}kers, D., 2008, arXiv:0805.3946.

\item[] Pereira, S., Pitrou, C.,  Uzan, J.-P., 2007, arXiv:0707.0736

\item[] Pullen, A. R., Kamionkowski, M., 2007, Phys. Rev. D 76, 103529

\item[] Prunet, S.,  Uzan, J.-P.,  Bernardeau, F., Brunier, T., 2005,
Phys. Rev. D71, 083508

\item[] Rakic, A., Rasanen, S., Schwarz,  D. J., 2006, MNRAS 
369, L27

\item[]  Ralston, J. P.,  Jain, P., 2004,  Int. J. Mod. Phys., 
{\bf{ D13}}, 1857 

\item[] Rodrigues, D. C., 2008, Phys. Rev. D77, 023534 

\item[] Saha, R., Jain, P. and Souradeep, T., 2006, ApJ Lett.,  645, L89

\item[] Samal, P. K., Saha, R., Jain, P. and Ralston, J. P., 2008,
MNRAS  385, 1718 

\item[]  Schwarz, D. J., Starkman, G. D., Huterer, D.,  Copi, C. J., 2004,
Phys. Rev. Lett.  93, 221301 

\item[] Slosar, A.,  Seljak, U., 2004, Phys. Rev. D70, 083002

\item[] Tegmark, M.,  de  Oliveira-Costa, A. and Hamilton, A., 2003,
Phys. Rev. D  68, 123523 

\item[] Vale, C., 2005, astro-ph/0509039

\item[] Weeks, J. R., 2004, astro-ph/0412231

\item[] Wiaux, Y., Vielva, P.,  Martinez-Gonzalez, E.,  Vandergheynst, P.,
2006, Phys. Rev. Lett. 96, 151303 

\item[] Yokoyama, S., Soda, J., 2008,  JCAP 0808, 005
\end{itemize}
\end{document}